# The Role of hybridization in $Na_xCoO_2$ and the Effect of Hydration

C.A. Marianetti[1], G. Kotliar[2], and G. Ceder[1,3]

12/19/03

[1]*Department of Materials Science and Engineering, Massachusetts Institute of Technology, Cambridge, Massachusetts 02139*
[2]*Center for Materials Theory, Department of Physics and Astronomy, Rutgers University, Piscataway, NJ 08854, USA*
[3]*Center for Materials Science and Engineering, Massachusetts Institute of Technology, Cambridge, Massachusetts 02139*

**Abstract**

Density functional theory (DFT) within the local density approximation (LDA) is used to understand the electronic properties of $Na_{1/3}CoO_2$ and $Na_{1/3}CoO_2(H_2O)_{4/3}$, which was recently found to be superconducting[1]. Comparing the LDA charge density of $CoO_2$ and the Na doped phases indicates that doping does not simply add electrons to the $t_{2g}$ states. In fact, the electron added in the $t_{2g}$ state is dressed by hole density in the $e_g$ state and electron density in the oxygen states via rehybridization. In order to fully understand this phenomenon, a simple extension of the Hubbard Hamiltonian is proposed and solved using the dynamical mean-field theory (DMFT). This simple model confirms that the rehybridization is driven by a competition between the on-site coulomb interaction and the hybridization. In addition, we find that the presence of $e_g$-oxygen hybridization effectively screens the low energy excitations. To address the role that water plays in creating the superconducting state, we compare the LDA band structure of $Na_{1/3}CoO_2$ and its hydrated counterpart. This demonstrates that hydration does cause the electronic structure to become more two-dimensional.

$Na_xCoO_2$ was originally investigated as a rechargeable battery material[2]. More recently, $Na_xCoO_2$ and its hydrated counterpart are receiving renewed interest due to the discovery of superconductivity[1], in addition to anomalous thermoelectric properties[3]. Understanding these properties requires a detailed understanding of the low energy Hamiltonian in this material. While there are many interesting parallels between the cuprates and $Na_xCoO_2$, the details of the electronic structure are quite different. In $Na_xCoO_2$, the Fermi level lies within the $t_{2g}$ states, with x=1 corresponding to a band insulator with filled $t_{2g}$ states and x=0 corresponding to one $t_{2g}$ hole per cobalt (possibly a Mott insulator). One of the useful steps forward in understanding the cuprates was the work of Zhang and Rice, in which they derived an effective one band model [4]. Initial studies of the low energy physics of $Na_xCoO_2$ have assumed a one band model[5-7], given that the octahedra in this structure are distorted and the $t_{2g}$ levels are split into a 1-fold $a_{1g}$ level and a 2-fold $e_g$' level (not to be confused with the usual $e_g$ levels) [8]. However, the potential influence of the $e_g$ orbitals and the oxygen orbitals on the low energy physics has not yet been addressed.

$Na_xCoO_2$ crystallizes in a layered structure, in which the layers are two-dimensional triangular lattices of a given species. The layers alternate as Na-O-Co-O and this pattern is repeated. Both the Na and the Co reside in sites which are coordinated by distorted oxygen octahedra. Insertion of water causes the oxygen layers to be pried apart, resulting in an even more two-dimensional-like structure [1]. Co exists in the low spin state due to the appreciable crystal field splitting. When x<1 Co is in a mixed valence state, $Co^{3+}/Co^{4+}$, yielding an average electronic configuration $t_{2g}^{6-x}$. However, the large overlap of the $e_g$ and the oxygen states creates strong hybridization and therefore an appreciable occupation of the $e_g$ states and hole density on the oxygen states.

Previous density functional theory (DFT) calculations addressing $Na_xCoO_2$ focused on the magnetic properties, in addition to rationalizing the transport properties in terms of the LDA density-of-states [9, 10]. However, hydration was not explicitly considered. In this letter, we present first-principles calculations of $Na_{1/3}CoO_2$ and $Na_{1/3}CoO_2(H_2O)_{4/3}$. We demonstrate that the naïve picture of adding electrons to the $t_{2g}$ states when doping $CoO_2$ with Na does not accurately describe the physics of $Na_xCoO_2$.

DFT calculations within the local density approximation (LDA) predict that Na doping adds electrons to the $t_{2g}$ states, in addition to hole density in the $e_g$ state and electron density in the oxygen states. Additionally, we demonstrate that the presence of the $e_g$-oxygen hybridization screens the low energy excitations of the system.

All LDA calculations were performed using the *Vienna Ab-initio Simulation Package* (VASP) [11, 12]. VASP solves the Kohn-Sham equations using projector augmented wave pseudopotentials [13, 14] and a plane wave basis set. A cut-off energy of 600 *eV* was chosen and k-point meshes of 6x6x3 were used for all cells. Experimental structural parameters were taken as a starting point for full structural relaxations [15, 16]. LDA calculations were performed for $Na_{1/3}CoO_2$ and $Na_{1/3}CoO_2(H_2O)_{4/3}$. The ordering of the Na atoms has been determined for both structures, and the ordering of the water molecules has been narrowed to a few possibilities [15, 16]. The Na atoms are four-fold coordinated by the water molecules, with two water molecules above and two below. However, there are six possible sites above and below the Na atom which may accommodate the water molecule. Two possibilities were suggested by Jorgensen et al[16] and we selected the more symmetric of the two configurations in which two opposing sites above the Na and the mirror image of the sites below the Na are occupied. Regardless of whether or not this particular ordering happens to be the ground state, the qualitative effect should be similar.

The LDA bands for $Na_{1/3}CoO_2$ and $Na_{1/3}CoO_2(H_2O)_{4/3}$ are presented in Figure 1. In both cases, a stable moment of 2/3 Bohr magnetons per formula unit is found. The respective bandstructures demonstrate that the $t_{2g}$ bands are similar, illustrating that hydration does not have any dramatic effect on the low energy physics. In both cases the dispersion in the *z* direction is small, which is a result of the two-dimensional nature of both the hydrated and non-hydrated compound. One notable difference is that the bands in the hydrated case display less splitting due to interlayer coupling at any given k-point. Given the supercell that we have chosen, only 9 $t_{2g}$ bands would be distinguishable at a generic k-point if the layers did *not* interact (ie. the unit cell contains 3 Co basis atoms per plane). Interlayer coupling will induce band splitting, and this is more pronounced for $Na_{1/3}CoO_2$ than $Na_{1/3}CoO_2(H_2O)_{4/3}$. This reflects the fact that the layers are more isolated in the hydrated case, and hence more two-dimensional, which is commonly believed to be an important ingredient of superconductivity in the cuprates[17]. It might be argued that the difference in band the structures are small, but it may still be relevant considering that the superconducting transition temperature is roughly 4 Kelvin.

Upon doping $CoO_2$ with Na, the Na will largely donate its electron at the Fermi energy. Therefore the changes in the ground state upon doping should qualitatively resemble the nature of the low energy excitations of the system. Given that the LDA DOS for $Na_xCoO_2$ shows the Fermi energy to lie within the $t_{2g}$ states, which only have a relatively small mixing with oxygen, one might expect that the electron addition state upon doping is some linear combination of $t_{2g}$ orbitals in addition to some small oxygen character. We shall demonstrate that this is not true, indicating that the LDA bands are behaving in an extremely non-rigid fashion. In order to accurately characterize the changes in the ground state upon doping $CoO_2$, it is useful to plot the charge density which has been added to the system. This can be done by subtracting the charge density of $CoO_2$ from that of $Na_{1/3}CoO_2$ calculated with the same structural parameters. This approach has been used frequently in studies of $Li_xCoO_2$ [18].

The difference in charge density between $CoO_2$ and $Na_{1/3}CoO_2$ is shown in Figure 2, which illustrates that the addition states are not simply $t_{2g}$ orbitals. In fact, electron density is added to the $t_{2g}$ states, hole density is added to the $e_g$ states, and electron density is added to the oxygen states. This can be explained intuitively as a multi-step process, and this hypothesis will be justified below. LDA predicts an appreciable occupation of the $e_g$ orbitals for $CoO_2$, roughly 1.6 electrons in a 1 Å sphere around Co, due to the large hybridization between the directly overlapping oxygen and $e_g$ states. When charge is added to the system, electron density in the $t_{2g}$ orbitals will increase causing an increase of the Co on–site coulomb repulsion. The on-site coulomb repulsion can be minimized by unmixing the oxygen and $e_g$ orbitals in order to decrease the occupation of the $e_g$ orbitals. Therefore, this rehybridization mechanism is a *competition* between the $e_g$-oxygen hybridization and the Co on-site coulomb interaction. Clearly, this logic can be inverted to explain the behavior of $Na_{1/3}CoO_2$ when Na is removed and holes are introduced into the system. In this case, holes are introduced into the $t_{2g}$ states, in addition to electrons in the $e_g$ states and holes in the oxygen states. The system increases the hybridization, and therefore the occupation of the $e_g$ orbitals, because there is less $t_{2g}$ density to interact with $e_g$ density. One can think of the rehybridization as a quasiparticle in which the particle (electron or hole) in the $t_{2g}$ states has been dressed by a rehybridization cloud. A more quantitative analysis of the effect can be given by integrating the change in charge density within a 1 Å spheres centered on each atom. The oxygens gain 0.73 electrons per unit cell, while Co only gains 0.12 electrons per unit cell due to the rehybridization mechanism. It should be noted that the Na potential clearly does have a notable effect on the rehybridization. The oxygen orbitals nearest the Na are preferentially occupied in order to more effectively screen the Na potential.

Hints of this rehybridization effect were first noticed computationally in a study of $Li_xTiS_2$ over 20 years ago[19]. The authors note that most of the net incoming charge ends up on the Sulfur as Li is added to the material. Studies of $Li_xCoO_2$ characterized this phenomena much more clearly[18, 20]. Wolverton and Zunger made a key step in understanding this phenomenon by suggesting that it is driven by a need to minimize the potential effect of the strong on-site coulomb interactions and that it is similar to what is observed for transition metal impurities in semiconductors[20]. We wish to emphasize that this phenomena is a competition between the on-site coulomb repulsion and the hybridization, and that the degree to which the rehybridization occurs will depend on the relative magnitudes of the parameters of the particular material at hand. This effect is likely to occur in many other transition metal oxides in which electrons or holes are being doped into the $t_{2g}$ states, such as $Na_xTiO_2$, $Li_xVO_2$. The results from X-ray Absorption experiments for $Li_xCoO_2$ and $Na_xCoO_2$ are consistent with the rehybridization mechanism[21-24]. The Co *2p* signal shows little variation as a function of doping, while the oxygen *1s* signal shows strong changes. This is in agreement with the observation that the rehybridization mechanism causes the average valence of the transition metal to remain relatively constant, while a net hole is formed on the oxygen.

While LDA gives an indication of the addition state, it cannot even qualitatively characterize the low energy excitations of the system when the electronic correlations are strong. If the rehybridization mechanism is indeed the result of a competition between hybridization and the on-site coulomb interaction, one should be able to capture this effect within a Hubbard-like model. The qualitative effects of the hybridization on the low energy excitations may then be explored within the modified Hubbard model. Consider the following model Hamiltonian:

$$H = \sum_{i,\sigma}\left[\varepsilon_p p_{i,\sigma}^\dagger p_{i,\sigma} + \varepsilon_e e_{i,\sigma}^\dagger e_{i,\sigma} + T_{p-e}\left(e_{i,\sigma}^\dagger p_{i,\sigma} + p_{i,\sigma}^\dagger e_{i,\sigma}\right)\right] + \sum_{i,j,\sigma} w_{i,j} t_{i,\sigma}^\dagger t_{j,\sigma}$$
$$+ U\sum_{i,\sigma}\left[e_{i,\sigma}^\dagger e_{i,\sigma} t_{i,\sigma}^\dagger t_{i,\sigma} + t_{i,\uparrow}^\dagger t_{i,\uparrow} t_{i,\downarrow}^\dagger t_{i,\downarrow} + e_{i,\uparrow}^\dagger e_{i,\uparrow} e_{i,\downarrow}^\dagger e_{i,\downarrow}\right] - \mu\sum_{i,\sigma}\left[p_{i,\sigma}^\dagger p_{i,\sigma} + e_{i,\sigma}^\dagger e_{i,\sigma} + t_{i,\sigma}^\dagger t_{i,\sigma}\right] \quad (1)$$

This model contains three orbitals ($p, t, e$) which can be analogously thought of as the oxygen $p$ orbitals, the transition metal $t_{2g}$ orbitals, and the transition metal $e_g$ orbitals. The respective annihilation operators are denoted as $p, e,$ and $t$. We only allow for hybridization between the $p$ orbital and the $e$ orbital as the oxygen orbitals hybridize much more strongly with the $e_g$ orbitals than the $t_{2g}$ for transition metals in an octahedral site. A coulomb repulsion $U$ is included for the $e$ and $t$ electrons, and for simplicity we only allow the $t$ electrons to hop. The hopping parameters for the $t$ electrons, $w_{ij}$, are chosen to yield a semicircular density of states of width 1 eV centered about zero. Additionally, we choose $\varepsilon_p$=-1.5 eV, $\varepsilon_e$=1 eV, $T_{p-e}$=2 eV, and $U$=3 eV. This model contains all the necessary ingredients to qualitatively demonstrate the rehybridization mechanism. It should be clear that if all the $e$ and $p$ terms were removed, we would be left with a simple one-band Hubbard model for the $t$ electrons.

The above model is not solvable in general, and we shall therefore employ the dynamical mean-field theory [25] (DMFT) as an approximation. Using a path integral representation of the partition function, the effective action for this model can be written in infinite dimensions as follows:

$$S_{eff} = \int_0^\beta\int_0^\beta d\tau d\tau' t^\dagger(\tau) G_t^o(\tau-\tau') t(\tau') + \int_0^\beta d\tau e^\dagger(\tau) G_e^o(\tau) e(\tau) +$$
$$U\int_0^\beta d\tau \sum_{i,\sigma}\left[e_{i,\sigma}^\dagger(\tau) e_{i,\sigma}(\tau) t_{i,\sigma}^\dagger(\tau) t_{i,\sigma}(\tau) + t_{i,\uparrow}^\dagger(\tau) t_{i,\uparrow}(\tau) t_{i,\downarrow}^\dagger(\tau) t_{i,\downarrow}(\tau) + e_{i,\uparrow}^\dagger(\tau) e_{i,\uparrow}(\tau) e_{i,\downarrow}^\dagger(\tau) e_{i,\downarrow}(\tau)\right]$$
(2)

where

$$G_e^o(\tau) = \left[\partial_\tau - \varepsilon_e + \mu - \frac{T_{p-e}^2}{\partial_\tau - \varepsilon_p + \mu}\right] \quad (3)$$

and $G_t^o$ is the bath function for the $t$ electrons which has to be determined self consistently using equation 4, where $D$ is the quarter bandwidth.

$$G_t^o(i\omega_n) = \left[ i\omega_n + \mu - D^2 G_t(i\omega_n) \right]^{-1} \quad (4)$$

The Grassman variables for the oxygen states have been integrated out. Due to the fact that we have assumed that the $e$ electrons have no intersite hopping, $G_e^o$ will remain unchanged. The above effective action corresponds to a two orbital AIM, and we are still left with a difficult many-body problem. In order to solve this, we follow the usual procedure of decoupling the quartic terms using a discreet Hubbard-Stratonavitch transformation[26], and evaluate the resulting summation using Hirsch-Fye quantum Monte-Carlo[25, 27].

First, we demonstrate how the occupation of the $t$ and $e$ orbitals change as the total density is varied (see Figure 3). When the $t$ orbital is nearly empty, the $e$ orbital has an occupation of roughly 0.45 electrons due to the strong hybridization with the oxygen. As the density increases and the $t$ orbital is filled, the $e$ orbital empties and the $p$ orbital is filled. Thus we clearly observe the rehybridization mechanism in this simple model. It does not appear to be as strong as in LDA calculation of $Na_xCoO_2$, where the gain in density in the $t_{2g}$ orbitals is largely cancelled by the loss in $e_g$ occupation. However, this is not totally surprising given the simple nature of the model we have chosen. Most notably, we only provided one $e$ orbital when there are two in the real material. This will decrease the initial occupation of the total $e_g$ density.

In order to explore the effect of the hybridization on the low energy excitations of the system, we have performed calculations at different values of $T_{p-e}$ while maintaining a half filled $t$ band (see Figure 4). The parameters chosen for this model yield a Mott insulator when the $t$ orbital is half filled, as seen by the gap in the spectral function. The gap is shown to decrease as $T_{p-e}$ is increased. This demonstrates that the hybridization among the $e$ and $p$ orbitals results in an effective screening of the $t$ electrons. This can be understood considering that increasing the hybridization will increase the occupation of the $e$ orbital, resulting in an increased repulsion with the $t$ electron and hence driving it to delocalize.

To conclude, we have shown that the states added to $CoO_2$ are not simply electrons in the $t_{2g}$ states, but rather an electron in the $t_{2g}$ state dressed by hole density in the $e_g$ state and electron density in the oxygen state. This rehybridization results from a competition between oxygen-$e_g$ hybridization and the Co on-site coulomb interactions. The LDA result is corroborated with a modified Hubbard model which we proposed and solved using DMFT. Our DMFT calculations indicate that the presence of $e_g$-oxygen hybridization effectively results in a screening of the low energy $t_{2g}$ excitations. This indicates that the oxygen states and $e_g$ states should be considered when deriving a low energy Hamiltonian for $Na_xCoO_2$ and its hydrated counterpart. We note that this behavior can be seen in several other similar materials, and that this is a rather general phenomenon. Comparing the LDA bands for $Na_{1/3}CoO_2$ and $Na_{1/3}CoO_2(H_2O)_{4/3}$ shows a reduction in the band splitting due to a decrease of the interlayer coupling. This indicates that hydration does result in a more two-dimensional electronic structure, and this fact may be related to why hydration is required to create the superconducting state.

This research was supported with funding from The Department of Energy, Basic Energy Science under contract DE-F602-96ER45571 and the National Partnership for Advanced Computing (NPACI).

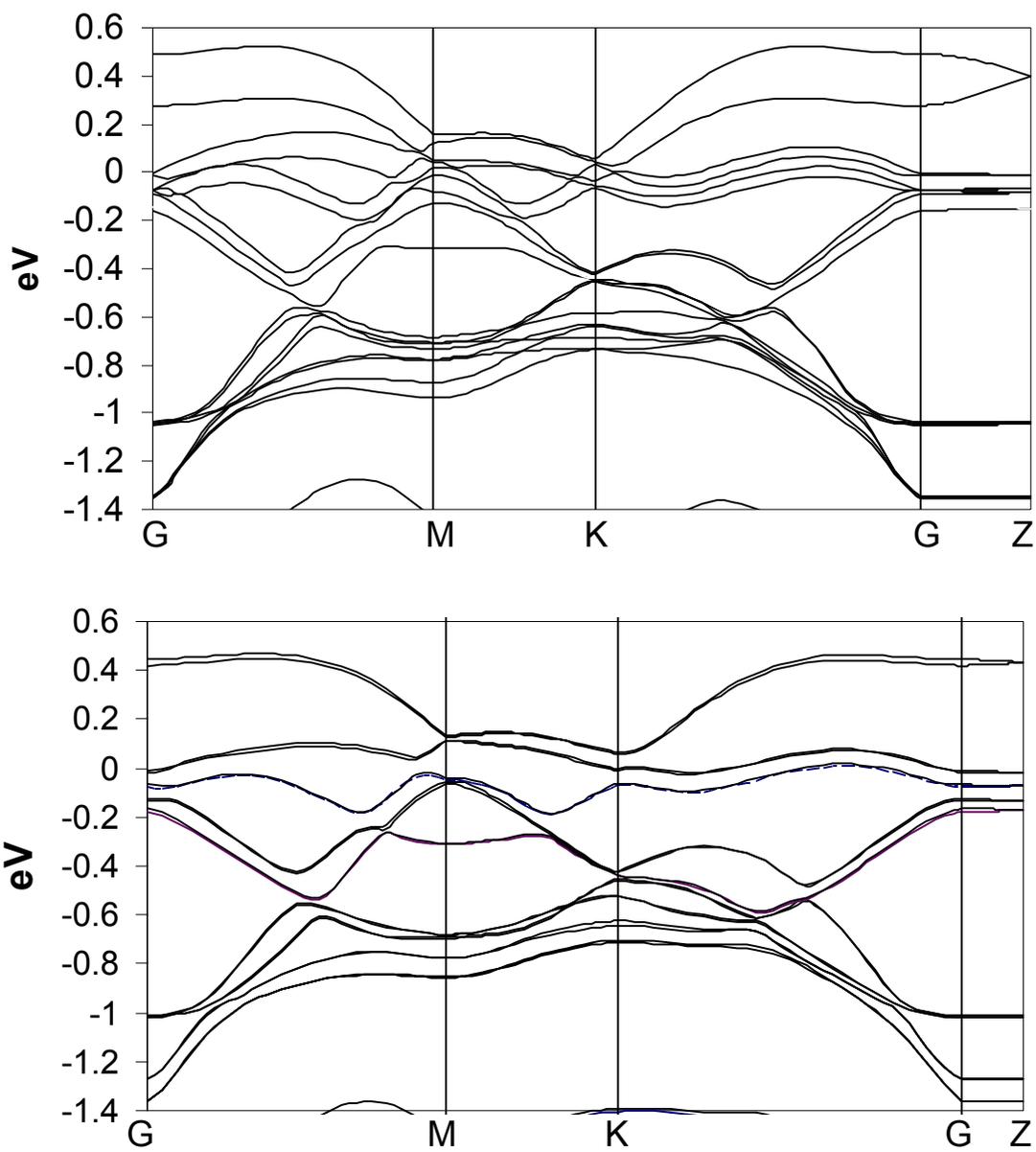

**Figure 1** LDA majority $t_{2g}$ bands for $Na_{1/3}CoO_2$ (top panel) and $Na_{1/3}CoO_2(H_2O)_{4/3}$ (bottom panel). The Fermi energy is at zero. G=(0,0,0), M=(1/2,0,0), K=(2/3,1/3,0), and Z=(0,0,1/2).

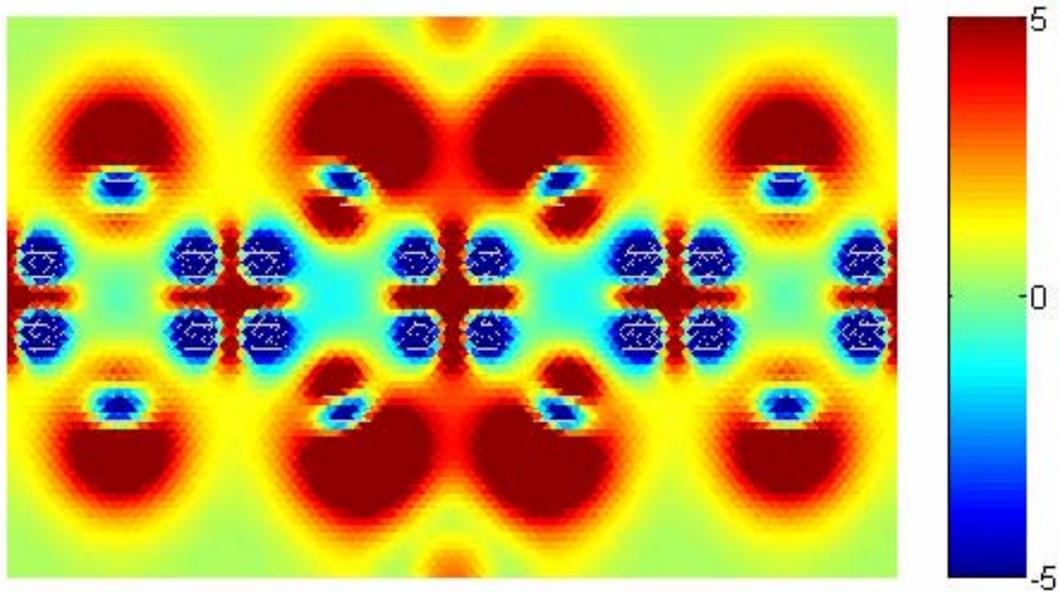

**Figure 2** The total charge density difference of $Na_{1/3}CoO_2 - CoO_2$. A plane which cuts through the Co-Oxygen octahedral plane was chosen. The horizontal rows of Na/vacancies, oxygen, and Co are clearly depicted (bottom and top edge of figure bisect Na/vacancy layer). Units are electrons/Å$^3$.

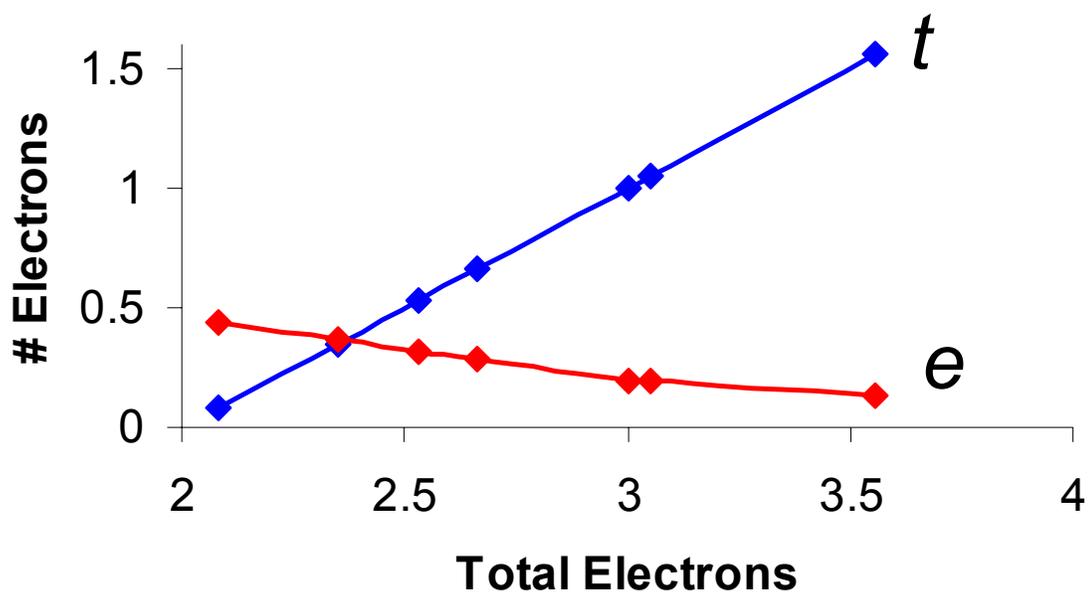

**Figure 3** Total occupation of the *e* and *t* orbitals as a function of the total density.

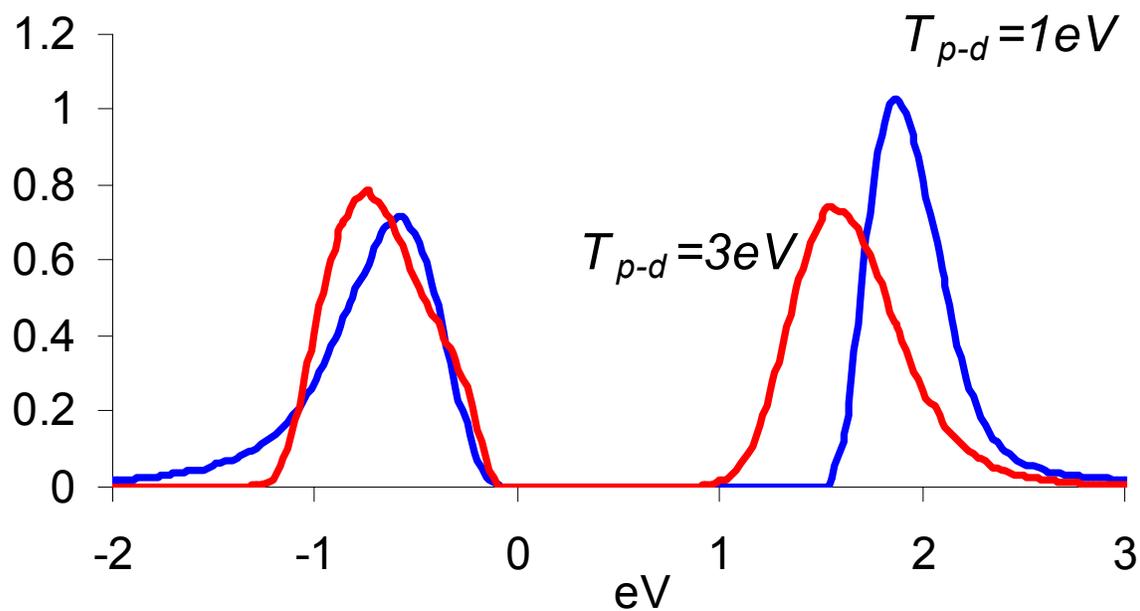

**Figure 4** Spectral density for the *t* electrons $\left[\frac{-1}{\pi}\mathrm{Im}G_t(\omega)\right]$